\documentclass[aps,prl,twocolumn,superscriptaddress,showpacs,longbibliography]{revtex4-2}
\usepackage[colorlinks=true,citecolor=blue,linkcolor=blue,breaklinks=true]{hyperref}
\usepackage[english]{babel}
\usepackage{amssymb,amsmath,amsfonts}
\usepackage{graphicx}

\usepackage{times}
\usepackage{xcolor}
\usepackage{subfigure}
\usepackage{setspace}
\usepackage{bm}% bold math

\usepackage{calc}
\usepackage{natbib} % elide or merge

\usepackage{booktabs} % for much better looking tables
\usepackage{array} % for better arrays (e.g. matrices) in maths

\begin{document}

%-----------------------------------------------------------------------------------------
\newcommand {\beq} {\begin{equation}}
\newcommand {\eeq} {\end{equation}}
\newcommand {\bqa} {\begin{eqnarray}}
\newcommand {\eqa} {\end{eqnarray}}
\newcommand {\ba} {\ensuremath{b^\dagger}}
\newcommand {\Ma} {\ensuremath{M^\dagger}}
\newcommand {\psia} {\ensuremath{\psi^\dagger}}
\newcommand {\psita} {\ensuremath{\tilde{\psi}^\dagger}}
\newcommand{\lp} {\ensuremath{{\lambda '}}}
\newcommand{\A} {\ensuremath{{\bf A}}}
\newcommand{\Q} {\ensuremath{{\bf Q}}}
\newcommand{\kk} {\ensuremath{{\bf k}}}
\newcommand{\qq} {\ensuremath{{\bf q}}}
\newcommand{\kp} {\ensuremath{{\bf k'}}}
\newcommand{\rr} {\ensuremath{{\bf r}}}
\newcommand{\rp} {\ensuremath{{\bf r'}}}
\newcommand {\ep} {\ensuremath{\epsilon}}
\newcommand{\nbr} {\ensuremath{\langle ij \rangle}}
\newcommand {\no} {\nonumber}
\newcommand{\up} {\ensuremath{\uparrow}}
\newcommand{\dn} {\ensuremath{\downarrow}}
\newcommand{\rcol} {\textcolor{red}}
%-----------------------------------------------------------------------------------------

\begin{abstract}
The scaling of entanglement entropy with subsystem size fails to distinguish between gapped and gapless ground state of a scalar field theory in $d>1$ dimensions. We show that the scaling of angular momentum resolved entanglement entropy $S_\ell$ with the subsystem radius $R$ can clearly distinguish between these states. For a massless theory with momentum cut-off $\Lambda$,  $S_\ell \sim \ln [\Lambda R/\ell]$ for $\Lambda R \gg \ell$, while $S_\ell \sim R^0$ for the massive theory. In contrast, for a free Fermi gas with Fermi wave vector $k_F$,  $S_\ell \sim \ln [k_F R]$ for $k_F R \gg \ell$. We show how this leads to an ``area-log'' scaling of total entanglement entropy of Fermions, while the extra factor of $\ell$ leads to a leading area law even for massless Bosons.
 \end{abstract}
%-----------------------------------------------------------------------------------------
\title{Signature of Criticality in Angular Momentum Resolved Entanglement of Scalar Fields in $d>1$}
\author{ Mrinal Kanti Sarkar} \email{mrinal.sarkar@tifr.res.in} \affiliation{Department of Theoretical Physics, Tata Institute of Fundamental
  Research, Mumbai 400005, India.}
\author{Saranyo Moitra} \email{smoitra@theory.tifr.res.in} \affiliation{Department of Theoretical Physics, Tata Institute of Fundamental
  Research, Mumbai 400005, India.}
\author{Rajdeep Sensarma}
 \affiliation{Department of Theoretical Physics, Tata Institute of Fundamental
 Research, Mumbai 400005, India.}

\date{\today}

\maketitle

The scaling of entanglement entropy with the size of the subsystem~\cite{LAFLORENCIE20161,Casini_2009, EEManybodyReview, EisertAreaLaws, jiang2012topological,KitaevPreskillTopo,Tarun_Ashvin_topo,LevinWen_Topo,PhysRevB.101.085136, vidal2003entanglement,Wilczek,CalabreseCardy,Hermelegapped} can provide strong indications about the nature of the underlying quantum state in a many-body system. For a generic thermal state, the entanglement entropy scales with the ``volume'' of the subsystem~\cite{Barthel2021,HusePal,ChakrabortyEnt2} (i.e. $S \sim R^d$ for a subsystem of linear size $R$ in $d$ spatial dimensions). Entanglement entropy of gapped ground states ~\cite{Latorre2006,CalabreseCardy,Hastings_2007} as well as excited states in many body localized systems~\cite{Abanin_EntanglementMBL,HusePal,Nandkishore:2014kca} scale with the ``area'' of the subsystem (i.e. $S \sim R^{d-1}$ ). Quantum scars~\cite{HOwen, Turner_scars} have a logarithmic scaling with subsystem size. 
Subleading scaling of entanglement entropy can also be used to identify gapped topological phases~\cite{KitaevPreskillTopo,Tarun_Ashvin_topo,LevinWen_Topo}. Dynamics of entanglement entropy can also reveal the nature of the quantum system~\cite{Abanin_EntanglementMBL,HoAbaninEntDyn,AhanaRajdeepBosons,moitra2020entanglement}.

Entanglement scaling in gapless ground states is more complicated.
In $d=1$, gapless ground states (e.g. free massless scalar~\cite{Calabrese_2009,Saravani_2014} fields as well as the free Fermi gas~\cite{CALLAN199455,HOLZHEY1994443,Goev2006}) show a logarithmic scaling of entanglement entropy with the subsystem size ($S \sim \ln ~R$), with a universal prefactor~\cite{Wilczek,CalabreseCardy,RyuTakayanagi2006,vidal2003entanglement} determined by the central charge of the $1+1$ D ~\footnote{We will use $d$ for spatial dimensions and D for space-time dimensions} %\cite{footnotedvD} 
conformal field theory~\cite{HOLZHEY1994443,CalabreseCardy,FradkinMoore}. For $d>1$, the answers vary: entanglement entropy of free fermions show an ``area-log'' scaling~\cite{Goev2006,Barthelchung,Swingle,FradkinMoore,Weifei2006, Wolf,PCalabrese_2012}, $S \sim R^{d-1} \ln R$. However, the ground state of massless scalar fields shows an area law entanglement scaling~\cite{Srednicki,Bombelli,SubirON,LOHMAYER2010222} in spite of being a gapless conformally invariant system. The leading order scaling of entanglement entropy does not show any signature of the gapless state, although subleading corrections can indicate the presence of criticality~\cite{LOHMAYER2010222,RyuTakayanagi2006}. This raises the question: why does the leading entanglement scaling of critical Bosons do not leave any signature in $d>1$?

In $d>1$, ground states of both massless and massive scalar fields show area law~\cite{Latorre2006,CalabreseCardy,Hastings_2007,Bombelli,Casini_2005,EisertAreaLaws} scaling of entanglement, with a non-universal prefactor which depends on the mass and the high energy regularization of the theory. One can ask: Is there any entanglement entropy related quantity whose leading order scaling with the subsystem size clearly distinguishes between the gapped and the gapless state?

In this paper, we answer both the questions raised above by considering the entanglement of a ``spherical'' subsystem $A$ of radius $R$ centred at the origin in the ground state of a free scalar field theory in 2+1 and 3+1 D with a momentum cut-off $\Lambda$. Due to the rotational invariance, the total entanglement entropy of the system $S$ is a sum of entropies in each angular momentum channel $\ell$, i.e. $S= \sum_\ell g_\ell S_{\ell}$, where $g_\ell =1$ in 2+1 D, and $g_\ell=(2\ell+1)$ in 3+1 D.  We show that:

(a) For massless scalar fields in both 2+1 and 3+1 D, $S_\ell \sim \frac{1}{6} \ln [\Lambda R/2\ell]$ for $1\ll \ell \ll  \Lambda R$. The logarithmic scaling of $S_\ell$ with $\Lambda R/\ell$ with a universal prefactor hints at an effective $1+1$ D CFT for each angular momentum channel. For massive theories,  $S_\ell \sim \textrm{const.}$ in this limit. Thus, the scaling of $S_\ell$ with subsystem size can clearly distinguish between gapless and gapped ground states for scalar fields in $d>1$. This is the key result of this paper. 

(b) For a massless scalar field, $S (\Lambda R)$ gets substantial contributions from $S_\ell$ with  $\ell < \ell_c$ with $2 \ell_c=\Lambda R$. The $\ln \Lambda R$ contribution from each $\ell$ channel leads to a leading area-log scaling, but this is exactly cancelled by the contribution of the $\ln \ell$ term. The subleading constant terms in $S_\ell$ then lead to the area law scaling of the total entanglement entropy with non-universal prefactors.

(c) For comparison, we also calculate $S_\ell$ for a free Fermi gas with Fermi wave vector $k_F$ in $d=2 $ and $d=3$. Here, $S_\ell$ is suppressed for $\ell \gg k_FR$ and $S_\ell \sim \frac{1}{6} \ln [k_FR]$ for $k_FR \gg \ell$. In this case, there is no cancellation: the number of channels scale as $(k_F R)^{d-1}$ and hence  $S\sim (k_FR)^{d-1} \ln [k_FR]$. For scalar fields, the extra scaling by $\ell$ in the logarithm for $S_\ell$ hides the signature of the gapless phase in the scaling of $S$, which becomes apparent when we look at scaling of individual $S_\ell$ s. Our results should also apply to leading order in an interacting $O(N)$ theory in the large $N$ limit~\cite{SubirON}, where they can be used to track the quantum phase transition in the system~\cite{SubirON,SubirWitczakKrempa}.

Beyond early experiments~\cite{Islam2015_expt}, entanglement entropy of 1 d Bosons have recently been measured in ultracold atomic systems~\cite{Tajik2023} using interference of identical systems. A similar arrangement with interference of 2 d pancakes can be used to construct $S_{\ell}$ and verify the scaling of $S_{\ell}$.

The action for free scalar field theory in $d$ dimensions is
\bqa
\mathcal{S}&=& \int d^d \rr ~\int dt~ \phi(\rr,t) \left[ -\partial_t^2+\nabla^2 -m^2\right] \phi(\rr,t)\\
\nonumber &=&  \int_0^\Lambda \frac{d^d \kk}{(2\pi)^d}~\int~\frac{d\omega}{2\pi} ~\phi(\kk,\omega) \left[ \omega^2-|\kk|^2 -m^2\right] \phi(-\kk,-\omega)
\eqa
where $m$ is the mass, and we have defined the theory with an ultraviolet momentum cut-off $\Lambda$ to get finite answers for entanglement entropy. $\Lambda^{-1}$ can be considered as a minimum grid size in real space for this system. The spectrum of the theory is given by $\omega_\kk=\sqrt{|\kk|^2+m^2}$, and $m=0$ corresponds to the gapless or critical theory. We will work with the ground state of this theory in this paper.

The von-Neumann entanglement entropy of free scalar fields can be calculated in terms of its correlation functions within the subsystem~\cite{Bombelli, Casini_2009}, i.e.
  \beq
  S=\text{Tr}_A\left[\frac{(\hat{M}+1)}{2} \ln\frac{(\hat{M}+1)}{2}-\frac{(\hat{M}-1)}{2} \ln\frac{(\hat{M}-1)}{2}\right],
  \eeq
  where the trace is over co-ordinates in $A$. Here
  \beq
  \nonumber \hat{M}^2(\rr,\rr')= \int_0^R d^d\rr_1 M^-(\rr,\rr_1)M^+(\rr_1,\rr'),
  \eeq
  with the equal time correlators of the fields and their conjugate momentas,
  \bqa
  M^-(\rr,\rr') &=& \langle \phi(\rr,t)\phi(\rr',t)\rangle = \int_0^\Lambda \frac{d^d \kk}{(2\pi)^d}~ \frac{e^{\mathbf{i} \kk \cdot (\rr-\rr')}}{\omega_\kk},\\
  \nonumber M^+(\rr,\rr') &=& \langle \dot{\phi}(\rr,t)\dot{\phi}(\rr',t)\rangle = \int_0^\Lambda \frac{d^d \kk}{(2\pi)^d}~ \omega_\kk~e^{\mathbf{i} \kk \cdot (\rr-\rr')}.
  \eqa
  \begin{figure*}[t] \includegraphics[width=\textwidth]{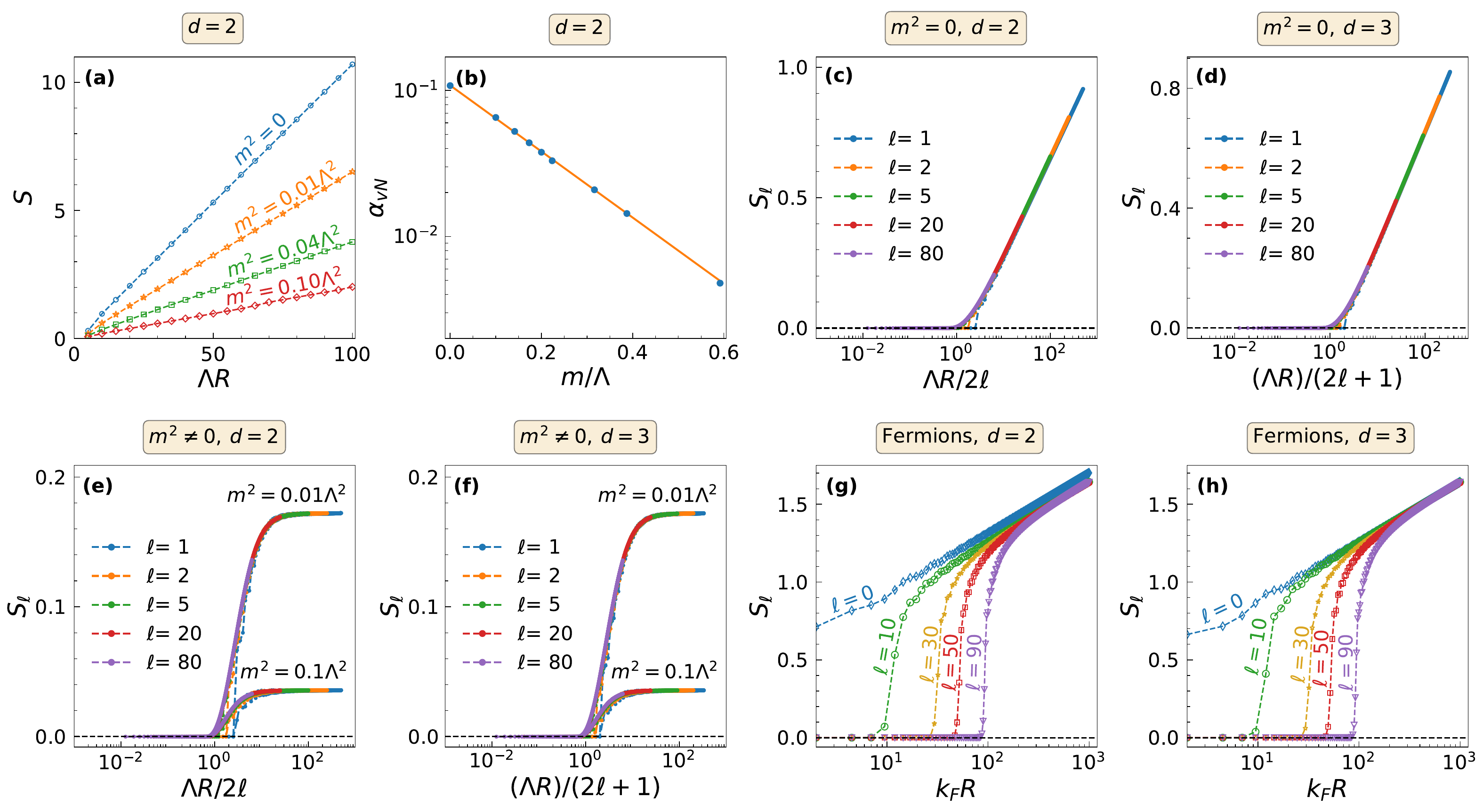}
  \caption{ Scaling of entanglement entropy with subsystem size for scalar field theory \textbf{(a)} -\textbf{(f)} and free Fermi gas \textbf{(g)}-\textbf{(h)}. \textbf{(a)} Area law scaling of $S= \alpha_{vN}(\Lambda R)$ + const. for a scalar field theory in 2+1 D for massless as well as massive systems.\textbf{(b)} $\alpha_{vN}$ as a function of $m/\Lambda$ showing exponential decrease. Note that $\alpha_{vN}$ is non-universal and depends on regularization scheme. \textbf{(c)}:: Logarithmic scaling of $S_\ell\sim \frac{1}{6} \ln ~[\Lambda R/2\ell]$  for a 2+1 D massless scalar field and \textbf{(d)} $S_\ell\sim \frac{1}{6} \ln ~[\Lambda R/(2\ell+1)]$  for a 3+1 D massless scalar field. Note the data collapse for different $\ell$ channels. \textbf{(e)}: Scaling of $S_\ell$ with $\Lambda R/2\ell$ for different $\ell$ channels for a massive 2+1 D scalar field theory with $m^2=0.01 \Lambda^2$ and $m^2=0.1\Lambda^2$. $S_\ell$ goes to a constant at large $\Lambda R$. The constant value decreases with increasing $m$. \textbf{(f)} Scaling of $S_\ell$ with $\Lambda R/(2\ell+1)$ for massive  scalar fields in 3+1 D. The behaviour is similar to that in \textbf{(e)}. \textbf{(g)} and \textbf{(h)}: Logarithmic scaling of $S_\ell$ with $k_FR$ for a Free fermi gas in $d=2$ \textbf{(g)} and $d=3$ \textbf{(h)}. Note the absence of the extra scale factor of $\ell$ in this case. }
  \label{Selllogscaling}
  \end{figure*}
  % \begin{figure}[t]
  % \includegraphics[width=\columnwidth]{Sell_ScalarField_2d_v2.pdf}
  % \caption{ Angular momentum resolved entanglement entropy $S_\ell$ of a $2+1$ $D$ scalar field theory for a ``spherical'' subsystem of size $R$ centred at the origin.\rcol{(a) Scaling of $S_\ell$ with  $\Lambda R$ and (b) Scaling of $S_\ell$ with  $\ln [\Lambda R/2\ell]$  for a massless scalar field. (a) shows that $S_\ell$ is substantial only for $2\ell < \Lambda R$ while (b) shows that $S_\ell \sim \ln [\Lambda R/2\ell]$ for $\Lambda R \gg 2\ell$. (c) Scaling of $S_\ell$ with  $\Lambda R$ and (d) Scaling of $S_\ell$ with  $\ln [\Lambda R/2\ell]$  for a massive scalar field with $m^2=0.1\Lambda^2$. (c) shows thresholding behaviour of $S_\ell$, although the threshold is not precisely at $\Lambda R/2$. (d) shows that $S_\ell$ grows logarithmically with $\Lambda R/\ell$ for intermediate ranges of $R$, before saturating to a constant value at the largest subsystem sizes. }}
  % \label{Sell2dscalar}
  % \end{figure}

  We discretize the radial coordinate and use the eigenvalues of $\hat{M}^2$ which are greater than $1$ ~\cite{Bombelli,Casini_2009,eisert2003introduction,PlenoEiserDreibig} to compute the entanglement entropy. In Fig.~\ref{Selllogscaling} (a), we plot $S$ as a function of $\Lambda R$ for 2+1 D scalar fields. We find the well-known area law scaling $S =\alpha_{vN} (\Lambda R)$ + const. for both massless and massive fields, with a non-universal $\alpha_{vN}$ which rapidly decreases with increasing $m$ [See Fig.~\ref{Selllogscaling} (b)]. One can use a different regularization scheme, where the radial fields are defined on a discrete lattice~\cite{Srednicki}. While the area law is robust, the value of $\alpha_{vN}$ will differ in the two schemes.
  
  For the rotationally invariant system, one can work with the angular momentum channels $\ell$. Since $M^{\pm}$ and hence $\hat{M}^2$ is block diagonal in $\ell$, this reduces the complexity to solving many one-dimensional problems in the radial co-ordinates. In this case, it is easy to see that $S=\sum_\ell g_\ell S_\ell$, where $g_\ell=1$ for $d=2$ and $g_\ell=2\ell+1$ in $d=3$ and the $\ell^{th}$ channel entanglement entropy
\beq
S_\ell = \text{Tr}_r \left[\frac{(\hat{M}_\ell+1)}{2}\ln\frac{(\hat{M}_\ell+1)}{2}-\frac{(\hat{M}_\ell-1)}{2}\ln\frac{( \hat{M}_\ell-1)}{2}\right]
\eeq
where $\text{Tr}_r$  is traced over radial coordinates in A. Here $\hat{M}^2_\ell(r,r') =\int_0^R dr_1~ r_1^{d-1} M^-_\ell(r,r_1)M^+_\ell(r_1,r')$, where
\beq
M^\pm_\ell(r,r_1) =\left(rr_1\right)^{-\nu}\int_0^\Lambda dk~ k ~\omega_k^{\pm 1}~ J_{\ell+\nu}(kr) J_{\ell+\nu}(kr_1) .
\eeq
with $\nu =(d-2)/2$.  The continuum operator $\hat{M}_\ell^2$ has eigenvalues $\lambda_n\geq1$~\cite{Bombelli,Casini_2009,eisert2003introduction,PlenoEiserDreibig}.  To compute $S_\ell$, we evaluate $\hat{M}^2_\ell(r,r')$ on a discrete set of radial points to construct a finite-dimensional matrix, but only consider $\lambda_n\geq 1$ to compute the required trace. We have also considered an alternate regularization where the radial fields are discretized on a lattice of finite length~\cite{Srednicki,Latorre2006,LOHMAYER2010222}. The leading order results for $S_\ell$ are same in both cases (see SM).
% \begin{figure}[t]
%   \includegraphics[width=\columnwidth]{Sell_ScalarField_3d_v2.pdf}
%   \caption{ Angular momentum resolved entanglement entropy $S_\ell$ of a $3+1$ D scalar field theory for a ``spherical'' subsystem of size $R$ centred at the origin. \rcol{(a) Scaling of $S_\ell$ with  $\Lambda R$ and (b) Scaling of $S_\ell$ with  $\ln [\Lambda R/(2\ell+1)]$  for a massless scalar field. (a) shows that $S_\ell$ is substantial only for $2\ell +1 < \Lambda R$ while (b) shows that $S_\ell \sim \ln [\Lambda R/(2\ell+1)]$ for $\Lambda R \gg 2\ell$. (c) Scaling of $S_\ell$ with  $\Lambda R$ and (d) Scaling of $S_\ell$ with  $\ln [\Lambda R/(2\ell+1)]$  for a massive scalar field with $m^2=0.1\Lambda^2$. (c) shows threshold behaviour of $S_\ell$, although the threshold is not precisely at $\Lambda R=2\ell+1$. (d) shows that $S_\ell$ grows logarithmically with $\Lambda R/(2\ell+1)$ for intermediate ranges of $R$, before saturating to a constant value at the largest subsystem sizes.}}
%   \label{Sell3dscalar}
% \end{figure}

We consider $S_\ell$ as a function of the dimensionless size of the subsystem $\Lambda R$ for a $2+1$ D scalar field theory. In Fig.~\ref{Selllogscaling} (c), we plot $S_\ell$ for a massless scalar field theory as a function of $\Lambda R/2\ell $ for different values of $\ell$. For $\Lambda R/2\ell \gg 1$, the curves for different $\ell$ (other than $\ell=0$) collapse on top of each other. The curve is linear when plotted on a logarithmic scale for $\Lambda R/2\ell$, with a slope which is numerically found to be $1/6$. Thus we find
\beq
 S_\ell \sim  \frac{1}{6} \ln  \left[\frac{\Lambda R}{2\ell}\right] 
\eeq
We have checked that this scaling is independent of regularization schemes (see SM). The logarithmic scaling with the universal prefactor of $1/6$ hints at an underlying $1+1 $ D CFT for the radial modes, with spatial coordinates scaled by the angular momentum quantum number $\ell$. However, while the equation of motion for the radial modes has a global scale invariance, they do not have conformal invariance in these coordinates due to the presence of the centrifugal barrier. Note that this is the leading scaling of $S_\ell$ in a spherical subsystem, and is distinct from the subleading logarithmic scaling of $S$ in a subsystem with sharp corners~\cite{Bueno_Mayers_Krempa_2015,Stoudenmire,Helmes_Wessel,Fabien_David_Nicolas_2015}. We can contrast this with the scaling of $S_\ell$ for massive scalar fields. In Fig.~\ref{Selllogscaling} (e), we plot $S_\ell$ as function of $\Lambda R/2\ell$ (on logarithmic scale) for different values of $\ell$ for $m^2/\Lambda^2 =0.01$ and $m^2/\Lambda^2 =0.1$. In both cases, we find that the $S_\ell$ curves collapse for $\Lambda R/2\ell \gg1$. The curves rise linearly for intermediate ranges of $\Lambda R$, mimicking the logarithmic behaviour of the critical theory, but settle down to a constant value for the largest subsystem sizes. Thus $S_\ell \sim R^0$ for a massive theory. The constant value, which decreases with increasing $m$, is non-universal and depends on the regularization scheme. This is consistent with the fact that the system appears critical until the subsystem size is larger than the correlation length $\xi \sim 1/m$ in the system.

A similar logarithmic scaling with the same prefactor, $S_\ell \sim (1/6) \ln [\Lambda R/(2\ell+1)]$ is seen for a massless  $3+1$ D scalar field theory [see Fig.~\ref{Selllogscaling} (d)], while the massive theories in 3+1 D also show $S_\ell \sim R^0$ [see Fig.~\ref{Selllogscaling} (f)]. In this case, a scaling with $\Lambda R/(2\ell +1)$ leads to a better data collapse. Thus we see that in contrast to the total entanglement entropy, the leading scaling of $S_\ell$ with subsystem size can be used to distinguish between the gapped and the critical ground state of the scalar field theory. This is the key result of this paper.

The leading order scaling of $S_\ell$ with subsystem size should also hold in an interacting theory, like a $O(N)$ theory~\cite{SubirON}, and can be used to detect a quantum phase transition in $2+1$ or $3+1$ D theories ~\cite{SubirON, SubirWitczakKrempa}. As one approaches the quantum phase transition in this theory from the disordered side, one would expect $S_\ell \sim \ln [\Lambda R/2\ell]$ at the critical point. Close to the critical point,  $S_\ell$ will show logarithmic growth till a scale $R/\ell \sim \xi$ before saturating. This correlation length $\xi$ would be diverging as one approaches the transition.

It is instructive to compare and contrast the scalar field theory with a system of conformally invariant non-interacting spinless Fermions in $d>1$. The Hamiltonian of the free Fermi gas  $H=\sum_{\kk} \frac{k^2 -k_F^2}{2m} c^\dagger_{\kk} c_{\kk}$, where $c^\dagger_{\kk}$ creates a Fermion with momentum $\kk$. The ground state is a spherical Fermi sea of radius $k_F$, where $k_F$ is related to the density $\rho$ by $\rho= (\Omega_d/d(2\pi)^d)~k_F^d$. The momentum distribution of the Fermions $n_\kk=\Theta(k_F-k)$ jumps from $1$ to $0$ at $k=k_F$. In this case, it is well known~\cite{Peschel_2009, Casini_2009,moitra2020entanglement} that the entanglement entropy of the system is given by
\beq
S = -\text{Tr}_A~\left[ \hat{C} \ln ~\hat{C} + (1-\hat{C}) \ln~(1-\hat{C})\right]
\eeq
where $C(\rr,\rp)= \langle c^\dagger_\rr c_\rp \rangle = \int d^d \kk~ n_\kk~ e^{i\kk\cdot (\rr-\rp)}$ is the one particle correlation function. When $A$ is a ``sphere'' of radius $R$ at  the origin, $S= \sum_\ell g_\ell S_\ell$, where
\beq
S_\ell=-\text{Tr}_r~\left[\hat{C_\ell} \ln ~\hat{C_\ell} + (1-\hat{C_\ell}) \ln~ (1-\hat{C_\ell})\right],
\eeq
with $C_\ell(r,r')= (rr')^{-\nu}\int_0^{k_F} ~dk~k~ J_{\ell+\nu}(kr)J_{\ell+\nu}(kr')$. In Fig.~\ref{Selllogscaling} (g) and (h), we plot the variation of $S_\ell$ with $k_FR$ for different values of $\ell$ for free Fermi gas in $d=2$ and $d=3$ respectively. In both cases, we find that for $k_FR \gg \ell$, $S_\ell$ scales logarithmically with $k_FR$. In fact, we find
\beq
S_\ell \sim \frac{1}{6}~\ln~[k_FR]
\eeq
i.e. the universal prefactor of the $1+1$ D CFT makes a reappearance. Note that, unlike the scalar field theory, there is no additional scaling by $\ell$ in this case. The entanglement of a free Fermi gas is understood in terms of radial chiral modes at each angle on a Fermi surface~\cite{Swingle}. Here we see that a similar argument holds for each angular momentum channel.

\begin{figure}[t]
  \includegraphics[width=\columnwidth]{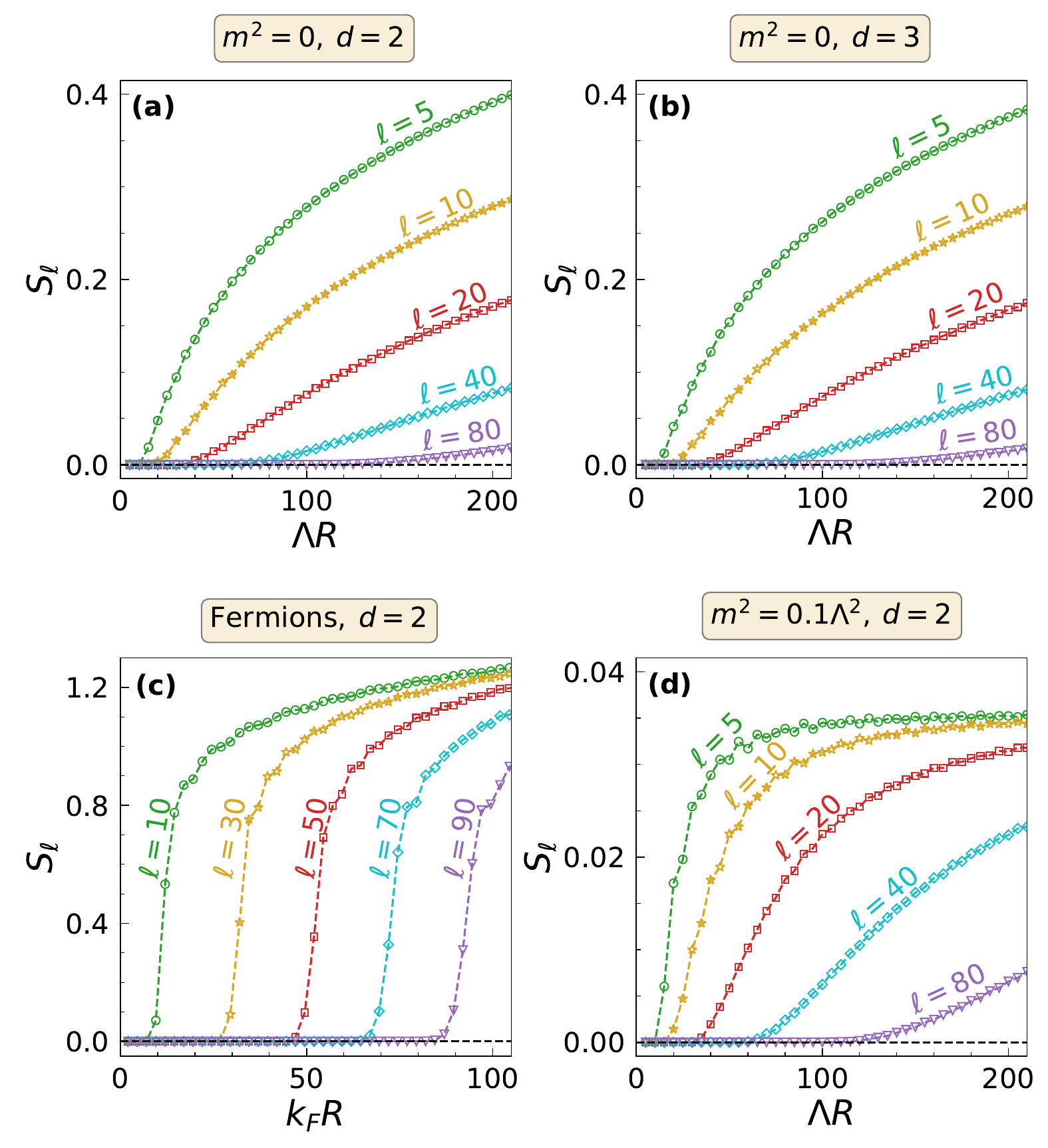}
  \caption{ Behaviour of $S_\ell$ with subsystem size for large $\ell$ or small $R$: Massless scalar fields in 2+1 D \textbf{(a)} and 3+1 D \textbf{(b)}: $S_\ell$ is strongly suppressed for $\Lambda R < 2\ell$ in 2+1 D and  $\Lambda R < 2\ell +1$ in 3+1 D. For a fixed $\Lambda R$, only $\ell <\ell_c$ contributes to $S$, where $2\ell_c \sim \Lambda R$. \textbf{(c)} $S_\ell$ for a free Fermi gas in $d=2$ is strongly suppressed for $k_F R < \ell$. For a fixed $k_FR$, only $\ell <\ell_c$ contributes to $S$, where $\ell_c \sim k_F R$. The logarithmic scaling of $S_\ell$ together with the cut-off $\ell_c$ explains the leading scaling of $S$ in \textbf{(a)}-\textbf{(c)}. \textbf{(d)} $S_\ell$ for a 2+1 D massive scalar field theory with $m^2=0.1 \Lambda^2$. The threshold of $\Lambda R$ increases with $\ell$, but $\ell_c/\Lambda R$ is not a universal number and varies weakly with $\ell$.}
  \label{Sellthreshold}
\end{figure}

We now turn our attention to the reverse question: if $S_\ell$ shows logarithmic scaling with subsystem size for both the massless scalar fields and free Fermi gas, why does the entanglement entropy show an area law for the bosons and an area-log law for the Fermions? To answer this question, we note that for the massless scalar fields in $d=2$, $S_\ell$ is strongly suppressed for $2\ell \gg \Lambda R$ [ See Fig.~\ref{Sellthreshold}(a)]. So, for a given $\Lambda R$, only angular momentum channels with $\ell < \ell_c =\Lambda R/2$ contribute. For 3+1 D critical scalar fields, a similar thresholding behaviour is seen with $2\ell_c+1= \Lambda R$ [Fig.~\ref{Sellthreshold}(b)]. The angular momentum $\ell$ corresponds to $\ell$ oscillations in the angular coordinate between $0$ and $2\pi$, with an angular period $\Delta \theta \sim 1/\ell$.  This corresponds to the largest transverse wavelength in the subsystem $\Delta R_t\sim R \Delta \theta \sim R/\ell$. For $\ell \gg \Lambda R$, $\Delta R_t $ is much smaller than the short distance scale $\Lambda^{-1}$, and these fast transverse oscillations wash out the contribution of these modes to $S$. A similar thresholding behaviour is also seen for the free Fermi gas with $\ell_c=k_FR$ [See Fig.~\ref{Sellthreshold}(c) for $d=2$ Fermi gas].

The area-log behaviour of $S$ for a free Fermi gas is now easy to understand: the number of $\ell$ modes scale as the area, $\sum_{\ell=0}^{k_FR} g_\ell \sim (k_FR)^{d-1}$ and each mode contributes $\sim \ln [k_F R]$. For the massless scalar field in 2+1 D, a similar counting argument gives $\sum_{\ell=1}^{\Lambda R/2} \ln [\Lambda R/2\ell] =\sum_{\ell=1}^{\Lambda R/2} \ln [\Lambda R/2]- \sum_{\ell=1}^{\Lambda R/2}  \ln \ell$. The first term gives an area-log scaling similar to the Fermions, but the second term gives $\ln (\Lambda R/2) ! \sim (\Lambda R/2) \ln (\Lambda R/2) -(\Lambda R/2)$ (see SM). The ``area-log'' terms cancel, leaving a leading order area law scaling of $S$. Similar cancellations occur for $d=3$ as well (see SM). Thus the extra factor of $\ell$ in the logarithmic scaling of $S_\ell$ leads to an exact cancellation of the universal terms, leaving a non-universal area law for entanglement entropy of critical Bosons in $d>1$. The scaling form also reproduces the fact that there are subleading logarithmic corrections to $S$ in 3+1 D, while such corrections are absent in 2+1 D (see SM)

In Fig.~\ref{Sellthreshold}(d), we plot the dependence of $S_\ell$ on $\Lambda R$ for a massive scalar field theory with $m^2=0.1\Lambda^2$ in 2+1 D. We again see thresholding, with the threshold in $\Lambda R$ increasing with $\ell$. However, $\ell_c/\Lambda R$ is not a universal number since the mass provides another length scale in the theory. Similar behaviour is also seen in 3+1 D massive theories. In this case, the area law can be explained by the number of angular momentum channels scaling with the area of the subsystem, with a constant contribution from each channel.

We have shown that unlike total entanglement entropy, scaling of angular momentum resolved entanglement entropy ($S_\ell$) of ground states of scalar field theories with subsystem size can clearly distinguish between massless and massive systems in $d>1$. $S_\ell$ shows logarithmic scaling with scaled variable $\Lambda R/\ell$ for critical theories and are independent of subsystem size for massive theories. In contrast, $S_\ell$ of a free Fermi gas grows logarithmically with $k_FR$ independent of the angular momentum quantum number. $S_\ell$ also shows a thresholding behaviour so that for a subsystem size  $R$, one needs to sum up to $\ell_c \sim \Lambda R/2$ for scalar fields and $\ell_c \sim k_FR$ for the Fermi gas. The extra factor of $\ell$ in the logarithmic scaling of $S_\ell$ cancels a leading ``area-log'' behaviour of total entanglement entropy for the critical scalar field theory, leaving an area law even for the critical case. The ``area-log'' scaling dominates for Fermions in the absence of the scale factor $\ell$. Our work thus sheds light on why the scaling of entanglement entropy in these two conformally invariant systems show dramatically different behaviour.

%%-----------------------------------------------------------------------------------------
\begin{acknowledgments}
MKS, SM and RS acknowledge the support of the Department of Atomic Energy, Government of India, for support under Project Identification No. RTI 4002.
\end{acknowledgments}

M.K.S. and S.M. contributed equally to this work.
%%-----------------------------------------------------------------------------------------
%-----------------------------------------------------------------------------------------
%\bibliographystyle{unsrt} 
\bibliography{Entanglement.bib}
\newpage
\onecolumngrid
\appendix
\section{Supplementary Material}

% -----------------------------------------------------------------------------------------

\section{Appendix A: Alternate Regularization and Robustness of Results}

It is well known~\cite{Srednicki,Bombelli,Latorre2006} that the entanglement entropy of the ground state of a scalar field theory (both massive and massless) in $d>1$ scales with the area of the subsystem, with a coefficient which is non-universal and depends on the regularization scheme. In the main text, we have shown that the angular momentum resolved entanglement, $S_\ell$ scales logarithmically with $\Lambda R/\ell$ for massless scalar fields. The natural question that arises is which of our results are dependent on the regularization scheme and which results are independent of it.

In the main text, we have used a regularization scheme with an ultraviolet momentum cut-off $\Lambda$ and defined the fields in the continuum. The calculation of the angular momentum resolved entanglement entropy $S_\ell$ involves the calculation of eigenvalues of the correlation function $\hat{M}^2_\ell(r,r^{\prime})$. While $\hat{M}^2$ is calculated as a continuum integral, we numerically evaluate it on a finite grid of equispaced radial points to construct a finite-dimensional matrix and consider its eigenvalues. While it is known that the continuum operator $\hat{M}^2_\ell$ has eigenvalues $\geq1$~\cite{Bombelli,Casini_2009,eisert2003introduction,PlenoEiserDreibig}, the finite-dimensional matrix has spurious eigenvalues $<1$. We neglect these spurious eigenvalues and use only the eigenvalues $\geq1$ for calculating the entanglement entropy. Our calculations do not change once the grid reaches $\Lambda^{-1}$. Note that the full system size is always set to infinity in our formulation.

One can use an alternate regularization scheme, as used by Srednicki in Ref.~\onlinecite{Srednicki,Latorre2006,LOHMAYER2010222}, where the radial Hamiltonian for each angular momentum channel $(\ell,\lbrace m_{\ell}\rbrace)$ in $d$ dimension becomes,
\begin{equation}%\tag{A1}\label{Hlm_eq1}
\begin{split}
H_{lm}=\frac{1}{2}\int_{0}^{\infty}dr \lbrace\Pi_{\ell,\lbrace m_{\ell}\rbrace}^2(r)+ r^{d-1}\left[\partial_r \left(\frac{\phi_{\ell,\lbrace m_{\ell}\rbrace}(r)}{r^{\frac{d-1}{2}}} \right)\right]^2+\left(\frac{\ell(\ell+d-2)}{r^2}+m^2\right)\phi_{\ell,\lbrace m_{\ell}\rbrace}^2(r) \rbrace 
\end{split}
\end{equation}
Here the fields $\phi_{\ell,\lbrace m_{\ell}\rbrace}$ and $\Pi_{\ell,\lbrace m_{\ell}\rbrace}$ are the projections of the scalar fields and their conjugate momenta into $d$ dimensional spherical Harmonics basis, satisfying $\left[\phi_{\ell,\lbrace m_{\ell}\rbrace}(r),\Pi_{\ell',\lbrace m_{\ell'}\rbrace}(r')\right]=i\delta_{\ell \ell'}\delta{m_{\ell}m_{\ell'}}\delta(r-r')$. 
The radial Hamiltonian is discretized on a finite lattice of N points with lattice constant a, which gives both UV and IR regularizations. The discrete Hamiltonian matrix is given by
\begin{equation}%\tag{A2}\label{Hlm_eqn2}
\begin{split}
H_{lm}=\frac{1}{2a}\sum_{i=1}^{N} \lbrace\Pi_{\ell,\lbrace m_{\ell},\rbrace,i}^2+ {(i+\frac{1}{2})}^{d-1}\left[ \frac{\phi_{\ell,\lbrace m_{\ell},\rbrace,i}}{{(i)}^{\frac{d-1}{2}}} -\frac{\phi_{\ell,\lbrace m_{\ell},\rbrace,i+1}}{{(i+1)}^{\frac{d-1}{2}}} \right]^2 +\left(\frac{\ell(\ell+d-2)}{r^2}+m^2\right)\phi_{\ell,\lbrace m_{\ell},\rbrace i}^2 \rbrace
\end{split}
\end{equation}
One can then map this to a Hamiltonian of harmonic oscillators and use it to calculate the entanglement entropy. If there are $N$ lattice points in the system and $n$ lattice points in the subsystem, one should consider the entanglement entropy in the limit $n/N \ll 1$ to get universal features.

In this section, we redo our calculations in the regularization scheme of Srednicki to show:
\begin{itemize}
	\item For massless fields,  $S_\ell \sim \frac{1}{6} \ln~[\Lambda R/ 2\ell]$ is independent of regularization scheme (including the prefactor of $1/6$). Subleading terms depend on the regularization scheme.
	
	\item For massive fields, $S_\ell \sim \text{const.}$ scaling is independent of the regularization scheme, but the constant value reached by $S_\ell$ for massive theories depends on the regularization scheme
	
	\item Hence the coefficient of the area law for $S$ depends on the regularization scheme
\end{itemize}

\begin{figure}[h!]
	\includegraphics[width=\columnwidth]{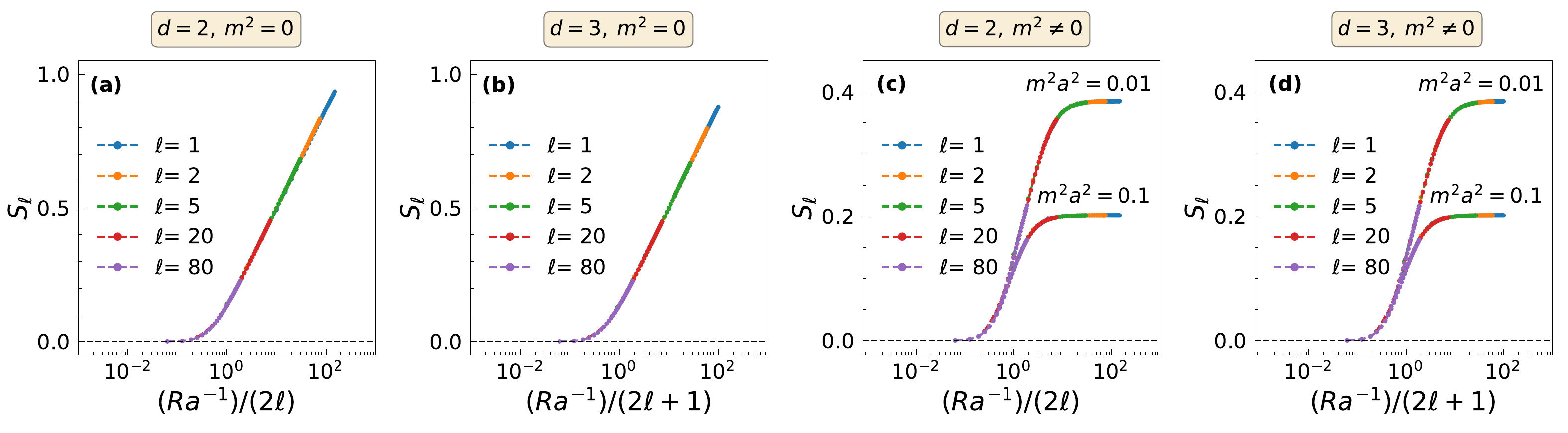}
	\caption{Angular momentum resolved entanglement entropy $S_{\ell}$ of 2+1 and 3+1 D scalar field theory using ``Sredinicki's regularization" on a finite lattice with lattice constant ``a" for a spherical subsystem of size R. \textbf{(a)} Scaling of $S_{\ell}$ with $[ Ra^{-1}/2\ell]$ for massless scalar field in $d=2$. \textbf{(b)} Scaling of $S_{\ell}$ with $[ Ra^{-1}/(2\ell+1)]$ for massless scalar field in $d=3$. \textbf{(c)} Scaling of $S_{\ell}$ with $[ Ra^{-1}/2\ell]$ of massive scalar fields in $d=2$ for $m^2a^2=0.01$ and $0.1$.\textbf{(d)} Scaling of $S_{\ell}$ with $[ Ra^{-1}/2\ell+1]$ of massive scalar fields in $d=3$ for $m^2a^2=0.01$ and $0.1$.}
	\label{Srednickiapproachslscalarfield}
\end{figure}

In Fig.~\ref{Srednickiapproachslscalarfield} (a) and (b), we plot $S_\ell$ for 2+1 and 3+1 D massless scalar fields, obtained in the alternate lattice regularization, as a function of $Ra^{-1}/(2\ell)$ and $Ra^{-1}/(2\ell+1)$ respectively. We find that $S_{\ell}\sim\frac{1}{6}\ln\left[\frac{Ra^{-1}}{2\ell}\right]$ in 2+1 D and $S_{\ell}\sim\frac{1}{6}\ln\left[\frac{Ra^{-1}}{2\ell+1}\right]$ in 3+1 D respectively. This matches with our results with $a =\Lambda^{-1}$. We have taken $N=1000$ and $n$ up to $300$ for these plots. In Fig.~\ref{Srednickiapproachslscalarfield} (c) and (d), we plot $S_\ell$ for massive scalar fields in $d=2$ and $d=3$ respectively. We see that $S_\ell$ saturates to a constant, which decreases with increasing $m$. However, the value of the constant is different in the two regularization schemes: e.g. in $d=2$, in our scheme, the constant value of $S_\ell \sim 0.17$ for $m^2=0.01 \Lambda^2$, while the Srednicki regularization scheme gives $S_\ell \sim 0.39$ for $m^2=0.01 a^{-2}$. This results in the coefficient of the area law for total entanglement entropy being regularization dependent.
Thus, our key results are robust to the vagaries of the regularization schemes used to calculate the entanglement entropy.

\section{Appendix B: Total entanglement entropy ($S$) from $S_{\ell}$: Angular momentum sum}

In the main text, we stated that the logarithmic scaling of $S_\ell$ with $\Lambda R/\ell$, together with a cut-off $\ell_c \sim \Lambda R$, leads to a cancellation of two ``area-log'' terms in the scaling of total entanglement entropy, leaving an area law with non-universal coefficients even for the massless scalar field. In this appendix we provide the details of these calculations.

For the gapless scalar field in 3+1 D, we have shown in the main text,
\begin{equation}
S_{\ell} \sim \frac{1}{6} \:\lbrace\ln (\frac{\Lambda R}{2\ell+1})- \ln B\rbrace\:\Theta[\ \ln(\frac{\Lambda R}{2\ell+1})-\ln B\ ].
\end{equation}
Here B is an $O(1)$ number and the $\Theta$  function gives an upper limit $\ell_c=(\ \frac{\Lambda R}{2B}-\frac{1}{2})\ $ of the angular momentum channels which adds a non-zero contribution to the entropy for a fixed system size $\Lambda R$. Note that, this approximate scaling function is not quantitatively accurate near the transition region where each of the $\ell$ channels starts rising to non-zero values from zero, but these inaccuracies give us subleading corrections. With this ansatz,
\begin{equation}\label{3d_saturation_eqn}
\begin{split}
S &\sim\frac{1}{6}\sum_{\ell=0}^{\ell_c}(2\ell+1)\:\lbrace \ln \left(\frac{\Lambda R}{2\ell+1}\right)-\ln B\:\rbrace=\frac{1}{6}[\ (\ell_c+1)^2\ln(2\ell_c+1)-\sum_{\ell=0}^{\ell_c}(2\ell+1)\ln(2\ell+1)]\ \\
&=\frac{1}{6}[\ \lbrace (\ell_c)^2\ln(2\ell_c)+\frac{\ell_c}{2} +(2\ell_c) \ln(2\ell_c)+\ln(\ell_c)+...\rbrace -\lbrace {(\ell_c)}^2\ln(2\ell_c)-\frac{{\ell_c}^2}{2} +(2\ell_c)\ln(2\ell_c)+\frac{11}{12}\ln(\ell_c)+... \rbrace]\ \\
&=\frac{1}{6}[\ \frac{\ell_c(\ell_c+1)}{2}+ \frac{1}{12} \ln(\ell_c)+...]\ \\ 
&\approx \frac{1}{48B^2}(\Lambda R)^2 + \frac{1}{144}\ln(\Lambda R)^2 + const.  
\end{split}%\tag{B2}
\end{equation}

Note the ``area-log'' term cancellation in the second line of~\eqref{3d_saturation_eqn}. For 2+1 D massless scalar fields, we have shown in the main text, $S_{\ell} \sim \frac{1}{6} \:\lbrace\ln(\frac{\Lambda R}{2\ell})- \ln B\rbrace\:\Theta[\ \ln(\frac{\Lambda R}{2\ell})-\ln B\ ]\ $ with $B \sim {\cal O}(1)$ for $\ell \neq 0$. Additionally, we have found that $S_{\ell=0}\sim \frac{1}{6} \ln \left( \frac{\Lambda R}{2B}\right)$. Summing over the angular momentum channels up to $\ell_c=(\frac{\Lambda R}{2B}) $ , we get
\begin{equation}\label{2d_saturation_eqn}
\begin{split}
S &\sim\frac{1}{6}\sum_{\ell=1}^{\ell_c}2\left[\ln\left(\frac{\Lambda R}{2\ell}\right)-\ln B\right]+ \frac{1}{6} \ln\left( \frac{\Lambda R}{2B}\right)
=\frac{1}{3}\sum_{\ell=1}^{\ell_c} \ln\left(\frac{\ell_c}{\ell} \right) + \frac{1}{6} \ln \left( \frac{\Lambda R}{2B}\right) \\
&=\frac{1}{3}[\ \ell_c \ln(\ell_c) - \ln(\ell_c!)]\ +\frac{1}{6}\ln \left( \ell_c\right)\approx \frac{1}{3}[\ \ell_c-\frac{1}{2}\ln(\ell_c)-\frac{1}{2}\ln(2\pi)+...]+\frac{1}{6} \ln \left(\ell_c \right) \\   
&= \frac{1}{6B}(\Lambda R) +const.
\end{split} %\tag{B3}
\end{equation}
In the second last line of~\eqref{2d_saturation_eqn}, we have used the Stirling approximation. Note that, the contribution of the $\ell=0$ mode exactly cancels the $\ln(\ell_c)$ term in the Stirling approximation. This explains the absence of logarithmic term in total entanglement entropy $S$ in 2+1 D in contrast to 3+1 D where we do get a logarithmic subleading term. Indeed, using this scaling function we get area law for 2+1 D, the entanglement entropy increase linearly with the system size of the subsystem, showing an area law. This simple scaling ansatz thus not only explains the leading area law but also explains the absence of subleading logarithmic terms in $d=2$, in contrast to the presence of such terms in $d=3$.
\begin{center}
\rule{3cm}{1pt}
\end{center}

%-----------------------------------------------------------------------------------------
%-----------------------------------------------------------------------------------------

\end{document}